\documentstyle[pra,aps,multicol,eqsecnum]{revtex}

\newcommand{\ket}[1]{|#1\rangle}
\newcommand{\bra}[1]{\langle#1|}
\newcommand{\mv}[1]{\langle#1\rangle}

\begin{document}
\title{Purification and correlated measurements of bipartite mixed states}
\author{Jan Bouda and Vladim{\'\i}r Bu{\v z}ek}
\address{
Faculty of Informatics, Masaryk University, Botanick\'a 68a,
602 00 Brno, Czech Republic  
\\
and
\\
 Research Center for Quantum Information,
Slovak Academy of Sciences,
D\'ubravsk\'a cesta 9, 842 28 Bratislava, Slovakia 
}
\maketitle
\begin{abstract}
We prove that all purifications of a non-factorable
state (i.e., the state which cannot be expressed in a form
$\rho_{AB}=\rho_A\otimes\rho_B$) are
 {\em entangled}.  We also show 
that for any  bipartite state there exists a pair of
measurements which are {\em correlated} 
on this state if and only if the state is
non-factorable.
\newline
{\bf PACS: 03.65.Ud, 03.65.Ta}
\end{abstract}

\begin{multicols}{2} 

\section{Introduction}
Quantum entanglement is one of the most important ingredient  of the 
paradigm of the quantum theory \cite{Peres1993,Alber2001}.
It plays the central role in quantum teleportation \cite{Bennett1993},
quantum dense coding \cite{Bennett1992}, quantum secret sharing
\cite{Hillery1999}, and other quantum information processes
\cite{Nielsen2000}.
 Quantum entanglement can be manipulated using
the entanglement swapping \cite{Zukowski1993,Bouda2001} and it can be
concentrated via quantum distillation techniques \cite{Nielsen2000}.

It is well known that 
pure entangled states  violate the
so-called Bell inequalities \cite{Bell1964}, which implies
that these states
 have nonlocal properties. This means that  pure entangled states
cannot be created locally. Moreover, for each bipartite pure state
there exists a pair of correlated measurements  \cite{Peres1993}
if and only if the state is entangled. 

In the case of mixed states the situation is more complex.
In 1989 Werner \cite{Werner1989} have introduced the following definition of
entanglement for mixed states: 
the bipartite mixed state is entangled if and only if it is
inseparable. In addition Werner has shown that
any separable state can be created exclusively
via local operations and classical communication (and hence
it doesn't have nonlocal properties).

In this Brief Report we will concentrate our attention 
on correlations in measurements performed on mixed entangled states.
The problem of correlations in measurements of two qubits 
 has been studied
by Englert \cite{Englert2001,Englert2000}.
Specifically, 
we will derive the
necessary and sufficient condition for existence of correlated measurements
on bipartite mixed states. 

In what follows we will
utilize the purification anzats as proposed by Uhlmann \cite{Uhlmann1976}
via which an impure state of a given quantum system 
can be purified with the help of ancillas. 
Our main motivation to study  purification of  mixed states is to determine
the relation between the entanglement present in purified states and
the existence of correlations in measurements performed on original 
bipartite mixed states. 
We will also study whether these correlations are related to non-locality
of purified states.

In Section \ref{sec_purification}
we introduce the notion of factorability and we derive the relation between
the factorability of bipartite density matrix $\rho$ and the entanglement
of any purification of $\rho$. In Section \ref{sec_correlation} we prove
that for any bipartite density matrix $\rho$ there exists
a pair of measurements which are correlated on $\rho$ if and only if $\rho$
is not factorable.

In order to unify the notation and terminology we define 
correlations in measurements and present two examples 
which clarify the problem we address.

Let $\rho_{AB}$ is a bipartite density matrix while
$\rho_A={\rm Tr}_B(\rho_{AB})$ and 
$\rho_B={\rm Tr}_A(\rho_{AB})$ are 
reduced density matrices of the subsystems $A$,
$B$, respectively. 
Let $E$, $F$ are measurements on the subsystems 
$A$, $B$, respectively, and
$\hat{E}$, $\hat{F}$ are the corresponding observables. 
Then the measurements $E$ and $F$ are
{\em correlated} on $\rho_{AB}$ if and only if
\begin{equation}
{\rm Tr}(\rho_{AB}\hat{E}\otimes\hat{F})\ne 
{\rm Tr}(\rho_A\hat{E}){\rm Tr}(\rho_B\hat{F}).
\end{equation}

{\em Example 1}: 
Let us consider 
two {\em correlated} sources $A,B$ emitting
spin-$\frac{1}{2}$ particles (qubits) such that with the probability
$\frac{1}{2}$ both sources simultaneously 
produce particles in the state $|0\rangle$ and
with the probability $\frac{1}{2}$ both particles are
simultaneously in the state $|1\rangle$. Hence, the sources
produce states 
$|00\rangle_{AB}$ or $|11\rangle_{AB}$ and the density matrix describing
this source is
\begin{equation}
\label{rho}
\rho_{AB}=\frac{1}{2}\left(|00\rangle_{AB}
\bra{00}+|11\rangle_{AB}\bra{11}\right).
\end{equation}
If we subject such pair of particles to 
orthogonal (projective) measurements in the bases
$\{ |0\rangle_A,|1\rangle_A\}$ 
and $\{\ket{0}_B,\ket{1}_B\}$, then the results of 
measurements of the state of particles $A$ and $B$ are the same. 
The reason is, that the pairs were produced in such 
a way that they are both in the same state.
In this case we can apply the formalism  of a micro-canonical ensemble 
since 
we have a set of pairs (of particles) denoted $p_1, p_2, \dots$ in pure 
states $\ket{\phi}_1, \ket{\phi}_2, \dots$, where each of the states
$\ket{\phi}_i$ is either $\ket{00}$ or $\ket{11}$. 
Results of the measurements are in this case two sequences of random variables
$a_1, a_2,\dots$
and $b_1, b_2, \dots$. 
Each pair of random variables
$a_i,b_i$ is not correlated. But the ensemble of the pairs (which
is described as a statistical mixture) exhibits correlatioms. 
As pointed out by Werner \cite{Werner1989}
these correlations have 
nothing to do with quantum non-locality and they are caused by 
classical correlations of the sources.

{\em Example 2}:
Now let us consider a source which repeatedly
produces three spin-$\frac{1}{2}$ particles $A,B,C$ in the 
Greenberger-Horn-Zeilinger (GHZ) state \cite{Nielsen2000}
\begin{equation}
\label{GHZ}
\ket{\phi}_{ABC}=
\frac{1}{\sqrt{2}}\left(\ket{000}_{ABC}+\ket{111}_{ABC}\right).
\end{equation}
Obviously, the reduced density
matrix $\rho_{AB}$ of particles $A,B$ is the same as
in the previous example described by Eq.~(\ref{rho}).
This means that  measurements 
in the bases $\{\ket{0}_A,\ket{1}_A\}$ and
$\{\ket{0}_B,\ket{1}_B\}$ yield the same results as in the previous example.
Nevertheless, in order to describe the present situation we have to 
 employ the macro-canonical formalism.
Results of the measurements are two random variables, which are
correlated,
and this correlation is caused by a quantum non-locality, which follows from
the fact that measurements $\{\ket{0}_A,\ket{1}_A\}$ and
$\{\ket{0}_B,\ket{1}_B\}$ performed on $\ket{\phi}_{ABC}$ are correlated due
to the present quantum entanglement.

In order to appretiate the relevance of these two examples we note
that in spite of the fact that the measurements performed on the
two systems generate the same experimental results their interpretation
might be totally different. 
The in-spite of this effect  is that although the
properties of separable states can be explained locally 
(i.e. without employing
entanglement), the actual physical reason behind these ``classical''
correlations  can be related to the quantum non--locality in 
preparation of the system.

\section{Purifications and factorability}
\label{sec_purification}

We start this section with the definitions of factorability,
separability and purification of density matrices:

The density matrix $\rho_{AB}$ is {\em factorable}
if it can be written in the form
\begin{equation}
\rho_{AB}=\rho_A\otimes\rho_B.
\end{equation}

The density matrix $\rho_{AB}$ is {\em separable}
if it can be written in the form
\begin{equation}
\rho_{AB}=\sum_i p_i\rho_A^{(i)}\otimes\rho_B^{(i)}.
\end{equation}

Let us consider a bipartite system AB in the state described by a 
density matrix
$\rho_{AB}$. Let $AB$ is a subsystem of some larger system
$ABC_1C_2$, which is in a pure state $\ket{\psi}$. 
Obviously, there is a whole class of states $|\psi\rangle$, 
which represent {\em
purifications}
of the density matrix $\rho_{AB}$, i.e. which fulfill the condition
\begin{equation}
{\rm Tr}_{C_1C_2}\left(\ket{\psi}\bra{\psi}\right)=\rho_{AB}.
\end{equation}
It is important to note that the purification of a given state 
$\rho_{AB}$ is not
unique. Firstly,  from the Schmidt decomposition \cite{Peres1993} it follows 
that  we can choose auxiliary systems 
$C_1$ and $C_2$ of an arbitrary dimension 
such that $\dim(C_1C_2)\geq\dim(AB)$.
Secondly, the purification is not unique even
when we fix dimensions of Hilbert spaces $C_1$ and $C_2$ because 
if $\ket{\psi}_{ABC_1C_2}$ is a {\em 
purification} of $\rho_{AB}$, then $U_{C_1C_2}\ket{\psi}_{ABC_1C_2}$
is also a purification of $\rho_{AB}$ for any unitary operator
$U_{C_1C_2}$ acting on $C_1C_2$.

{\bf Theorem 1.}
{\em Let $\rho_{AB}$ is a non-factorable density matrix.
Then any purification $\ket{\psi}_{ABC_1C_2}$ of $\rho_{AB}$ is entangled
in a sense that it cannot be written in the factorized form
\begin{equation}
\ket{\psi}_{ABC_1C_2}=\ket{\psi_1}_{AC_1}\otimes\ket{\psi_2}_{BC_2}.
\end{equation}
Conversely, if all purifications of $\rho_{AB}$ are entangled, then
$\rho_{AB}$ is non-factorable.}

In order to prove this theorem let us 
suppose that there is a purification
$\ket{\psi}_{ABC_1C_2}=\ket{\psi_1}_{AC_1}\otimes\ket{\psi_2}_{BC_2}$ of
$\rho_{AB}$. From the definition of purification it holds that
\begin{equation}
{\rm Tr}_{C_1C_2}(\ket{\psi}_{ABC_1C_2}\bra{\psi})=\rho_{AB}.
\end{equation}
However, from the definition of the partial trace we have
\begin{eqnarray}
{\rm Tr}_{C_1C_2}(\ket{\psi_1}_{AC_1}\ket{\psi_2}_{BC_2}\,
{{}_{AC_1}\langle \psi_1|}
{{}_{BC_2}\langle \psi_2|}
\nonumber
\\
={\rm Tr}_{C_1}(\ket{\psi_1}_{AC_1}\bra{\psi_1})\otimes
{\rm Tr}_{C_2}(\ket{\psi_2}_{BC_2}\bra{\psi_2})
\nonumber
\\
=\rho_A'\otimes\rho_B',
\end{eqnarray}
which is in a contradiction with the fact, that $\rho_{AB}$ is a non-factorable
density matrix.

In order to 
prove the second implication we will prove the following:
If $\rho_{AB}$ is factorable, then there exists a purification of
$\rho_{AB}$ which is not entangled. In fact, we will prove a stronger
statement by restricting the dimension of the purification.
Let $\dim({\cal H}_{AB})=n$. Then there exists a purification of $\rho_{AB}$
of dimension $n^2$, which is not entangled. It is well known, that there
exist purifications $\ket{\phi_1}_{AC_1}$ of $\rho_A$ and
$\ket{\phi_2}_{BC_2}$ of $\rho_B$ such that 
$\dim({\cal H}_A)=\dim({\cal H}_{C_1})$
and $\dim({\cal H}_B)=\dim({\cal H}_{C_2})$. Then $\ket{\psi}_{ABC_1C_2}=
\ket{\phi_1}_{AC_1}\otimes\ket{\phi_2}_{BC_2}$ is a purification of
$\rho_{AB}$ of the desired dimension, which is not entangled.

The Theorem 1 can be easily generalized for n--partite systems
in the following way: 
Let $\rho_{A_1\dots A_n}$ is a density matrix, which is not factorable in the
sense that it cannot be written as $\rho_{A_1\dots A_n}=\rho_{A_1}\otimes\dots
\otimes\rho_{A_n}$.
Then any purification $\ket{\psi}_{A_1\dots A_n C_1\dots C_n}$ of
$\rho_{A_1\dots A_n}$ is entangled
in the sense that it cannot be written in the form
\begin{equation}
\ket{\psi}_{A_1\dots A_n C_1\dots C_n}=\ket{\psi_1}_{A_1C_1}\otimes\dots
\otimes\ket{\psi_n}_{A_nC_n}.
\end{equation}
Conversely, if each purification of $\rho_{A_1\dots A_n}$ is entangled, then
$\rho_{A_1\dots A_n}$ is not factorable.

From above it follows that 
if $\rho_{AB}$ is a non-factorable state and $\rho_{ABC_1C_2}$ an arbitrary
mixed state such that ${\rm Tr}_{C_1C_2}(\rho_{ABC_1C_2})=\rho_{AB}$, then
$\rho_{ABC_1C_2}$ is not factorable in the sense that it cannot be written as
$\rho_{ABC_1C_2}=\rho_{AC_1}\otimes\rho_{BC_2}$.
This follows from Theorem 1, because each purification of
$\rho_{ABC_1C_2}$ is also a purification of $\rho_{AB}$ and thus it is
entangled.

It is also straightforward to show that a factorable 
 density matrix $\rho_{AB}$ has both
entangled and unentangled purifications.
Specifically, 
from the factorability we have $\rho_{AB}=\rho_A\otimes\rho_B$. Let
$\ket{\psi}_{AC_1}$ is a purification of $\rho_A$ and $\ket{\phi}_{BC_2}$
is a purification of $\rho_B$. Then $\ket{\psi}_{AC_1}\otimes\ket{\phi}_{BC_2}$
is a purification of $\rho_{AB}$. Let
$\ket{\omega}=U_{C_1C_2}\left(\ket{\psi}_{AC_1}\otimes\ket{\phi}_{BC_2}\right)$,
where $U_{C_1C_2}$ is a unitary operator acting on $C_1C_2$. Clearly
$\ket{\omega}$ is a purification of $\rho_{AB}$ for any $U_{C_1C_2}$ and
moreover there is a $U_{C_1C_2}$ such that $\ket{\omega}$ is entangled.

We conclude the present section by the following observation: 
If  a system
$AB$, which is a part of a larger system $ABC_1C_2$, is in a non-factorable
state $\rho_{AB}$, then it must be a part of a larger system which is 
{\em entangled}.
In other words, when we have a non-factorable system, then any larger
system (in a pure state) containing this system is entangled.
Moreover, 
for each purification $\ket{\psi}$ of $\rho_{AB}$ 
no 
unitary $U_{C_1C_2}$ operation can be found 
such that $U_{C_1C_2}\ket{\psi}$ is unentangled.
Hence, the non-factorability of $\ket{\psi}$ is not caused by the correlation
between $C_1$ and $C_2$. The most interesting fact is that all previous
statements hold regardless if $\rho_{AB}$ is separable or not.

\section{Correlations in measurements}
\label{sec_correlation}

{\bf Theorem 2.}
Let $\rho_{AB}$ is a non-factorable density matrix. Then there exists a pair
of orthogonal measurements represented by observables $E$ and $F$
(measured on ${\cal H}_A$ and ${\cal H}_B$, respectively), which are correlated
on $\rho_{AB}$.

{\em Proof}.
In order to prove the theorem we will use 
the negated implication. That is, 
let $\rho_{AB}$ is a density matrix such that
any two orthogonal measurements $E$ and $F$ performed on $\rho_{AB}$
are uncorrelated. Then $\rho_{AB}=\rho_A\otimes\rho_B$ is factorable.

A result of a measurement can be represented as a random variable.
Therefore the results of the measurement 
are uncorrelated iff the corresponding
random variables are uncorrelated, i.e., the covariance $C(E,F)$ fulfills 
the condition
$C(E,F)=0$.
Let $\rho_A={\rm Tr}_B(\rho_{AB})$ and $\rho_B={\rm Tr}_A(\rho_{AB})$, 
then the covariance of uncorrelated measurements fulfills the condition
\begin{equation}
C(E,F)=\mv{E\otimes F}_{\rho_{AB}}-\mv{E}_{\rho_A}\mv{F}_{\rho_B}=0, 
\end{equation}
from which it follows that
\begin{eqnarray}
\label{uncorrel}
{\rm Tr}(E\otimes F\rho_{AB})&=&{\rm Tr}(E\rho_A){\rm Tr}(F\rho_B)
\nonumber
\\
&=& 
{\rm Tr}(E\otimes F\rho_A\otimes\rho_B).
\end{eqnarray}
We want to show, that this identity implies $\rho_{AB}=\rho_A\otimes\rho_B$
and hence that $\rho_{AB}$ is factorable.

The condition (\ref{uncorrel}) holds for any two Hermitian operators
$E$ and $F$. Let us  choose some fixed basis $\{\ket{\phi_i}_{AB}\}_i$ on
${\cal H}_A\otimes {\cal H}_B$.  We will show that
\begin{equation}
\label{elementeq}
\forall i,j:\ (\rho_{AB})_{ij}=(\rho_A\otimes\rho_B)_{ij}.
\end{equation}
Because Eq.~(\ref{uncorrel}) holds for any two Hermitian operators
$E_i$ and $F_i$ it also holds that
\begin{equation}
\sum_i\alpha_i{\rm Tr}(E_i\otimes F_i\rho_{AB})=
\sum_i\alpha_i{\rm Tr}(E_i\otimes F_i\rho_A\otimes\rho_B)
\end{equation}
for any $\alpha_i\in C$ and hence
\begin{equation}
{\rm Tr}\left(\sum_i\alpha_iE_i\otimes F_i\rho_{AB}\right)=
{\rm Tr}\left(\sum_i\alpha_iE_i\otimes F_i\rho_A\otimes\rho_B\right).
\end{equation}
To prove Eq.~(\ref{elementeq}) it is enough to show that
\begin{equation}
{\rm Tr}(A\rho_{AB})={\rm Tr}(A\rho_A\otimes\rho_B)
\end{equation}
for any matrix $A$ such that
\begin{equation}
A_{ij}=1 \text{ for fixed }i,j\text{ and }A_{xy}=0\text{ otherwise}.
\end{equation}
However, an arbitrary matrix on ${\cal H}_A\otimes 
{\cal H}_B$ can be expressed as
\begin{equation}
\sum_i\alpha_iE_i\otimes F_i,
\end{equation}
where $E_i$ and $F_i$ are Hermitian matrices and $\alpha_i$
is an arbitrary complex number. This completes the proof.

The remaining part of this problem is trivial. 
When $\rho_{AB}=\rho_A\otimes\rho_B$
($\rho_{AB}$ is factorable), then the systems $A$ and $B$ are not
correlated which follows from Theorem 2.

\section{Conclusion}
We proved that any purification of a non-factorable state is
always {\em entangled}. 
This means that any system which  contains a non-factorable 
subsystem is also non-factorable. 
Moreover, we described  purifications of factorable states 
and we proved that for any bipartite density matrix $\rho$ there exists
a pair of measurements which are correlated on $\rho$ if and only if $\rho$
is non-factorable. Taking into account the fact  that  any purification of a
non-factorable  state  is entangled we conclude  
that these correlations have their origin in quantum non-locality.

This can be interpreted as an alternative approach to  Werner's  
explanation of the origin of correlations in measurements on separable (but
non-factorable) states. Our approach supplements the original work of Werner
\cite{Werner1989}. Specifically, we 
showed that  correlations on bipartite mixed state exists 
if and only if the state is non-factorable. 
These correlations can be explained
locally (see  Werner \cite{Werner1989}) 
when the state is separable, or they can be
explained via quantum entanglement of purified states (see 
Section II).

\acknowledgements
This work was supported 
by the European Union  projects EQUIP and QUBITS under the contracts
IST-1999-11053 and IST-1999-13021, respectively.
We also acknowledge a  support of the grant GACR 201/01/0413.
We thank M\'ario Ziman for many helpful discussions.

\end{multicols} 
\end{document}